\documentclass{elsart}
\usepackage{ifthen,graphics}
\usepackage{color}
\usepackage{cite}
\usepackage{epsfig}
\catcode`@=11 
\newdimen\z@ \z@=0pt 
\newskip\z@skip \z@skip=0pt plus0pt minus0pt
\def\m@th{\mathsurround=\z@}
\def\ialign{\everycr{}\tabskip\z@skip\halign} 
\def\eqalign#1{\null\,\vcenter{\openup\jot\m@th
    \ialign{\strut\hfil$\displaystyle{##}$&$\displaystyle{{}##}$\hfil
      \crcr#1\crcr}}\,}
\def\bye{\bibliographystyle{elsart-num}
  \bibliography{paper}
  \end{document}
}

\def \gev  {{\rm \,Ge\kern-0.125em V}}
\def \mev  {{\rm \,Me\kern-0.125em V}}
\def \kev  {{\rm \,ke\kern-0.125em V}}
\def \ev   {{\rm \,e\kern-0.125em V}}

\newcommand{\pvec}{\ensuremath{\mathbf{\vec{p}}}}

\newcommand{\dnc}{\ensuremath{d_{NC}}}
\newcommand{\Estar}{\ensuremath{E^*_{\gamma}}}
\newcommand{\qstar}{\ensuremath{\theta^*_{\gamma}}}

\newcommand{\dtTCA}{\ensuremath{\mathrm{d}_{\mathrm{tc}\bot}}}

\newcommand{\dc}{\ensuremath{d_\mathrm{c}}}
\newcommand{\kl}{\mbox{$K_{L}$}}
\newcommand{\ks}{\mbox{$K_{S}$}}
\newcommand{\ke}{\mbox{$K^0_{e3}$}}

\newcommand{\keg}{\mbox{$K^0_{e3\gamma}$}}
\newcommand{\kmgf}{\mbox{$K^0_{\mu 3(\gamma)}$}}
\newcommand{\kegf}{\mbox{$K^0_{e3(\gamma)}$}}

\newcommand{\klpeng}{\mbox{$K_{L}\to\pi^{\pm}e^{\mp}\nu(\gamma)$}}

\newcommand{\klpmn}{\mbox{$K_{L}\to\pi\mu\nu$}}

\newcommand{\kspp}{\mbox{$K_{S}\to\pi^+\pi^- $}}
\newcommand{\klppp}{\mbox{$K_{L}\to\pi^+\pi^-\pi^0$}}

\newcommand{\tca}{\mbox{TCA}}
\newcommand{\tof}{\mbox{TOF}}
\newcommand{\kloe}{\mbox{KLOE}}

\newcommand{\br}{\mbox{BR}}

\newcommand{\ie}{{\em i.e.}}
\newcommand{\vs}{{\em vs}}
\newcommand{\wrt}{{\em w.r.t.}}

\newcommand{\aff}[2]{Dipartimento di Fisica dell'Universit\`a #1 e Sezione INFN, #2, Italy.}
\newcommand{\affd}[1]{Dipartimento di Fisica dell'Universit\`a e Sezione INFN, #1, Italy.}
\begin{document}
  \begin{frontmatter} 
    \title{A Study of the Radiative \klpeng\ Decay and 
     Possible Osservation of Direct Photon Emission with the \kloe\ Detector}
%
\collab{The KLOE Collaboration}
\author[Na]{F.~Ambrosino},
\author[Frascati]{A.~Antonelli},
\author[Frascati]{M.~Antonelli},
\author[Frascati]{F.~Archilli},
\author[Roma3]{C.~Bacci},
\author[Karlsruhe]{P.~Beltrame},
\author[Frascati]{G.~Bencivenni},
\author[Frascati]{S.~Bertolucci},
\author[Roma1]{C.~Bini},
\author[Frascati]{C.~Bloise},
\author[Roma3]{S.~Bocchetta},
\author[Roma1]{V.~Bocci},
\author[Frascati]{F.~Bossi},
\author[Roma3]{P.~Branchini},
\author[Roma1]{R.~Caloi},
\author[Frascati]{P.~Campana},
\author[Frascati]{G.~Capon},
\author[Na]{T.~Capussela},
\author[Roma3]{F.~Ceradini},
\author[Frascati]{S.~Chi},
\author[Na]{G.~Chiefari},
\author[Frascati]{P.~Ciambrone},
\author[Frascati]{E.~De~Lucia},
\author[Roma1]{A.~De~Santis},
\author[Frascati]{P.~De~Simone},
\author[Roma1]{G.~De~Zorzi},
\author[Karlsruhe]{A.~Denig},
\author[Roma1]{A.~Di~Domenico},
\author[Na]{C.~Di~Donato},
\author[Pisa]{S.~Di~Falco},
\author[Roma3]{B.~Di~Micco},
\author[Na]{A.~Doria},
\author[Frascati]{M.~Dreucci\corauthref{cor1}},
\author[Frascati]{G.~Felici},
\author[Frascati]{A.~Ferrari},
\author[Frascati]{M.~L.~Ferrer},
\author[Frascati]{G.~Finocchiaro},
\author[Roma1]{S.~Fiore},
\author[Frascati]{C.~Forti},
\author[Roma1]{P.~Franzini},
\author[Frascati]{C.~Gatti},
\author[Roma1]{P.~Gauzzi},
\author[Frascati]{S.~Giovannella},
\author[Lecce]{E.~Gorini},
\author[Roma3]{E.~Graziani},
\author[Pisa]{M.~Incagli},
\author[Karlsruhe]{W.~Kluge},
\author[Moscow]{V.~Kulikov},
\author[Roma1]{F.~Lacava},
\author[Frascati]{G.~Lanfranchi},
\author[Frascati,StonyBrook]{J.~Lee-Franzini},
\author[Karlsruhe]{D.~Leone},
\author[Frascati]{M.~Martini},
\author[Na]{P.~Massarotti},
\author[Frascati]{W.~Mei},
\author[Na]{S.~Meola},
\author[Frascati]{S.~Miscetti},
\author[Frascati]{M.~Moulson},
\author[Frascati]{S.~M\"uller},
\author[Frascati]{F.~Murtas},
\author[Na]{M.~Napolitano},
\author[Roma3]{F.~Nguyen},
\author[Frascati]{M.~Palutan},
\author[Roma1]{E.~Pasqualucci},
\author[Roma3]{A.~Passeri},
\author[Frascati,Energ]{V.~Patera},
\author[Na]{F.~Perfetto},
\author[Lecce]{M.~Primavera},
\author[Frascati]{P.~Santangelo},
\author[Na]{G.~Saracino},
\author[Frascati]{B.~Sciascia},
\author[Frascati,Energ]{A.~Sciubba},
\author[Pisa]{F.~Scuri},
\author[Frascati]{I.~Sfiligoi},
\author[Frascati,Novo]{A.~Sibidanov},
\author[Frascati]{T.~Spadaro},
\author[Roma1]{M.~Testa},
\author[Roma3]{L.~Tortora},
\author[Roma1]{P.~Valente},
\author[Karlsruhe]{B.~Valeriani},
\author[Frascati]{G.~Venanzoni},
\author[Frascati]{R.Versaci},
\author[Frascati,Beijing]{G.~Xu}
\\
\address[Frascati]{Laboratori Nazionali di Frascati dell'INFN, 
Frascati, Italy.}
\address[Karlsruhe]{Institut f\"ur Experimentelle Kernphysik, 
Universit\"at Karlsruhe, Germany.}
\address[Lecce]{\affd{Lecce}}
\address[Na]{Dipartimento di Scienze Fisiche dell'Universit\`a 
``Federico II'' e Sezione INFN,
Napoli, Italy}
\address[Pisa]{\affd{Pisa}}
\address[Energ]{Dipartimento di Energetica dell'Universit\`a 
``La Sapienza'', Roma, Italy.}
\address[Roma1]{\aff{``La Sapienza''}{Roma}}
\address[Roma3]{\aff{``Roma Tre''}{Roma}}
\address[StonyBrook]{Physics Department, State University of New 
York at Stony Brook, USA.}
\address[Beijing]{Permanent address: Institute of High Energy 
Physics of Academica Sinica,  Beijing, China.}
\address[Novo]{Permanent address: Budker Institute of Nuclear Physics, Novosibirsk, Russia.}
\address[Moscow]{Permanent address: Institute for Theoretical 
and Experimental Physics, Moscow, Russia.}
\vspace{.5cm}
\begin{flushleft}
\corauth[cor1]{cor1}{\small $^1$ Corresponding author: Marco Dreucci
INFN - LNF, Casella postale 13, 00044 Frascati (Roma), 
Italy; tel. +39-06-94032696, e-mail marco.dreucci@lnf.infn.it}
\end{flushleft}
\setcounter{page}{1}        
\begin{abstract}
We present the measurement of the ratio
R = $\frac{\Gamma(\keg;\Estar>30\mev,\qstar>20^\circ)}{\Gamma(\kegf)}$ and a first measurement 
of the direct emission contribution for the same process.
We use 328 pb$^{-1}$ of data collected at \kloe\ in 2001 and 2002, corresponding to about 3 
million of \kegf\ events and about 9 thousand \keg\ radiative events. Our result is 
R= $(924 \pm 23_{stat} \pm 16_{syst})\times10^{-5}$ for the branching ratio and  
$\langle X\rangle=-2.3\pm 1.3_{stat}\pm 1.4_{syst}$ for the parameter describing
direct emission.
\\
\vspace{.5cm}
{\it key words:}~direct emission 
\end{abstract} 
\end{frontmatter}

\section{Introduction}
\label{sec:intro}
The study of radiative \kl\ decays offers the possibility to obtain informations on the kaon 
structure and the opportunity to test theories describing hadron interactions and decays,
like chiral perturbation theory (ChPT). Two different contribute to the photon 
emission, inner bremsstrahlung (IB) and direct emission (DE). 
DE is radiation from intermediate hadronic states and is thus sensitive to hadron structure.
In \ke\, DE is 1\% or less of IB which diverges both at \Estar$\to$0 and \qstar$\to$0 
(photon angle \wrt\ lepton). We therefore exclude small angle and energy photon.
To compare our result with other measurements we only retain events with \Estar$>30\mev$ and 
\qstar$>20^{\circ}$ \cite{doncel}. 
We define R as
\begin{equation}
  R = \frac{\Gamma(\keg;\Estar>30\mev,\qstar>20^\circ)}{\Gamma(\kegf)}
  \label{eq:ratio}
\end{equation}
Predictions for R ranges between 0.95$\times10^{-2}$ and 0.97
$\times10^{-2}$ ~\cite{kubis}. Recent measurements of R from NA48 and KTeV 
~\cite{NA48:R,KTeV:R} are in marginal disagreement between each other, so that new measurements are 
welcome.
Following the authors of Ref.\cite{kubis}, in ChPT the structure-dependent (SD) terms are 
characterized by six amplitude \{$V_i$, $A_i$\}, which in the one-loop approximation are real 
function, almost constant over phase space. 
In particular, all relevant SD terms have a similar and simple photon energy spectrum, with a 
maximum around \Estar $\sim$ 100 \mev. This suggests to decompose the photon spectrum in the 
following manner  
\begin{equation}
  \frac{d\Gamma}{d\Estar} \simeq \frac{d\Gamma_{IB}}{d\Estar} + \langle X\rangle f(\Estar)
  \label{eq:x}
\end{equation}
in which the different SD contributions are summarized in the so-called {\it{distortion function}},
$f(\Estar)$, which represents the deviation from the pure inner bremmstrahlung. 
All the information on the structure-dependent terms is contained in the effective 
strength, $\langle X \rangle$, that multiplies $f(\Estar)$. ChPT calculation at $\mathcal{O}(p^6)$ 
order from Ref~\cite{kubis} gives
\begin{equation}
  \langle X\rangle = -1.2 \pm 0.4  
\end{equation}
A first attemp to measure the DE contribution was performed in 2001 by KTeV collaboration
\cite{KTeV:DE}, but the uncertainties due to their working hypothesis were too large to infer
definitive conclusions on the $\langle X \rangle$ parameter.\\
In our analysis we can isolate DE from IB only because we use both the energy spectrum and 
the angular distribution of the radiated photon.

\section{The KLOE detector}
The KLOE detector consists of a large cylindrical drift chamber (DC), surrounded by a 
fine grained lead-scintillating fiber electromagnetic calorimeter (EMC). A superconducting 
coil around the calorimeter provides an axial magnetic field of about 0.52 T. 

The drift chamber ~\cite{KLOE:DC}, 4~m in diameter and 3.3~m long, is made of 58 concentric rings 
of drift cells arranged in a stereo geometry. It is filled with a $He,iC_4H_{10}$ mixture. 
The spatial resolutions are $\sigma_{xy}\simeq$0.15~mm and $\sigma_z\simeq$2~mm. 
The transverse momentum resolution is $\sigma_{p_{\perp}}/p_{\perp}\simeq 0.4\%$. 
Two-track vertices are reconstructed with a spatial resolution of $\sim$ 3~mm. 

The calorimeter ~\cite{KLOE:EmC} is divided into a barrel and two endcaps. It covers about 
98\% of the solid angle. It is segmented in depth in five layers, about $3X_0$ each.
The barrel is divided in 24 sectors, $5\times 12$ calorimeter cells each, read out by
photomultipliers at both ends to measure the arrival time of particles and to reconstruct
the space cordinates. Cells close in time and space are grouped into calorimeter clusters. 
The energy and time resolutions are $\sigma_E/E=$ $5.7\%/\sqrt{E~{\rm (GeV)}} $  and 
$\sigma_T=$ $54~{\rm ps}~\sqrt{E~{\rm (GeV)}} \oplus100~{\rm ps}$, respectively.
The spatial resolution are $\sigma_{xy}\simeq$1.3~cm and 
$\sigma_z\simeq $1~cm~$/\sqrt{E{\rm (GeV)}}$.  

The KLOE trigger~\cite{KLOE:trig} uses calorimeter and chamber information. For this 
analysis, only the calorimeter signals are used. 
Two energy deposits above threshold ($E>50$ MeV for the barrel and  $E>150$ MeV for 
the endcaps) are required. Recognition and rejection of cosmic-ray events is also 
performed at the trigger level. Events with two energy deposits above a 30 MeV 
threshold in the outermost calorimeter plane are rejected.
The 328 pb$^{-1}$ of data used in this analysis, taken in 2001 and 2002, 
are divided in 14  periods of about 25 pb$^{-1}$/period.
For each data period we have a corresponding period simulated with Monte Carlo (MC) 
with about the same statistic.

\section{Monte Carlo Generator}
In the \kloe\ MC only radiation from inner bremsstrahlung is described, so we need 
a Monte Carlo to describe the photon spectrum from direct emission. For this
purpose we use a Monte Carlo generator ({\it{Kubis}} generator) based on the code provided by one 
of the autors of Ref.~\cite{kubis}, which implements their $\mathcal{O}(p^6)$ calculation. 
We generate DE events folded with the \kloe\ reconstruction MC.
The accuracy of \kloe\ Monte Carlo in describing the photon spectrum from IB is at level of 
$\sim$1-2\% (an appropriate level for many \kloe\ measurements).
In particular, the \kloe\ MC generator avoids the problem of the infinite value for the total 
decay width for a single photon emission by re-summing, in the limit of soft photon 
energy, the probabilities for multiple photon emission to all order in $\alpha$~\cite{mcgatti}.
Unfortunately, this accuracy level is of the same order of DE contribution, which is $\sim$1\% of IB 
one. From a point of view of the measurement of R, this could introduce only a $\sim$1\% error. 
On the other side, a fit counting procedure based on a 1\%-biased IB distribution could introduce 
up to $\sim$100\% error counting in \keg\ from DE events. 
For this reason in this analysis we use the {\it{Kubis}} Monte Carlo generator also to describe the 
photon spectrum from IB.

\section{Analysis}
\begin{figure}[ht]
   \centering
   \psfig{figure=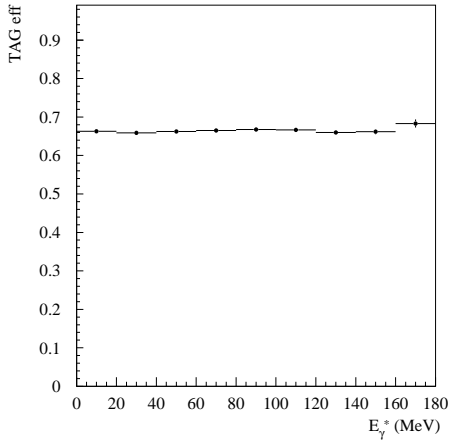,width=6.0cm}
   \caption{Tagging efficiency from Monte Carlo simulation as a function of photon energy. 
     Full statistic}
   \label{fig:efftag}
 \end{figure}

The criteria used to select \kegf\ events, briefly summarized below, are the same described in 
Ref.~\cite{kloe:ff}.
Candidate \kl\ events are tagged by the presence of a \kspp\ decay. Fig.~\ref{fig:efftag} shows that 
the tagging efficiency, about 66\%, is almost independent of the photon energy.
The \kl\ is searched along the direction of its momentum ({\it{tagging line}}), reconstructed 
from the decay \kspp.
\kegf\ events are then selected using appropriate kinematical variables of the decay and electron
identification by time of flight (\tof).
We have about $\sim$3 million of \kegf\ events with a contamination of 7$\times 10^{-3}$, 
mainly due to \kmgf\ events.

To select \keg\ events we search for a {\it{photon cluster}}, \ie, a cluster in the calorimeter 
not assigned to any track.
The arrival time of each photon gives an independent determination of the position of the \kl\ 
vertex, $\vec{X}_N$, the so-called neutral vertex (NV). The method is fully described in Refs.
~\cite{KLOE:EmC} and~\cite{KLOE:offline}. The position of the \kl\ vertex 
is assumed to be along the \kl\ line of flight.
We require that the distance, \dnc, between the position $\vec{X}_N$ of the neutral vertex and the
position $\vec{X}_C$ of the \kl\ vertex determined with tracks, to be within 8$\sigma$.
In case of more than one photon candidate, we choose the closest to the \kl\ charged vertex.
To evaluate the photon energy we use the charged track momenta and the photon cluster position, 
$\vec{X}_{\rm{clu}}$. By solving equation \ref{eq:eneg} below in the hypothesis of neutrino 
zero mass, we extract the photon energy with a resolution of $\sim$ 1\mev. This resolution is 
a factor $\sim$ 10 better than that obtained using the energy deposit information of the calorimeter.
\begin{equation}
  p_{\nu} = p_K - p_{\pi} - p_e - p_{\gamma}  ~~~;
  ~~~  \pvec_{\gamma} = E_{\gamma}~\frac{\vec{X}_{\rm{clu}}-\vec{X}_N}{|\vec{X}_{\rm{clu}}-\vec{X}_N|} 
  \label{eq:eneg}
\end{equation}
In this equation $p_\nu$, $p_K$, $p_{\pi}$, $p_e$ and $p_{\gamma}$ are the neutrino, kaon, pion, 
electron and photon momentum, respectively.

\begin{figure}[ht]
  \centering
  \begin{tabular}{cc}
    \psfig{figure=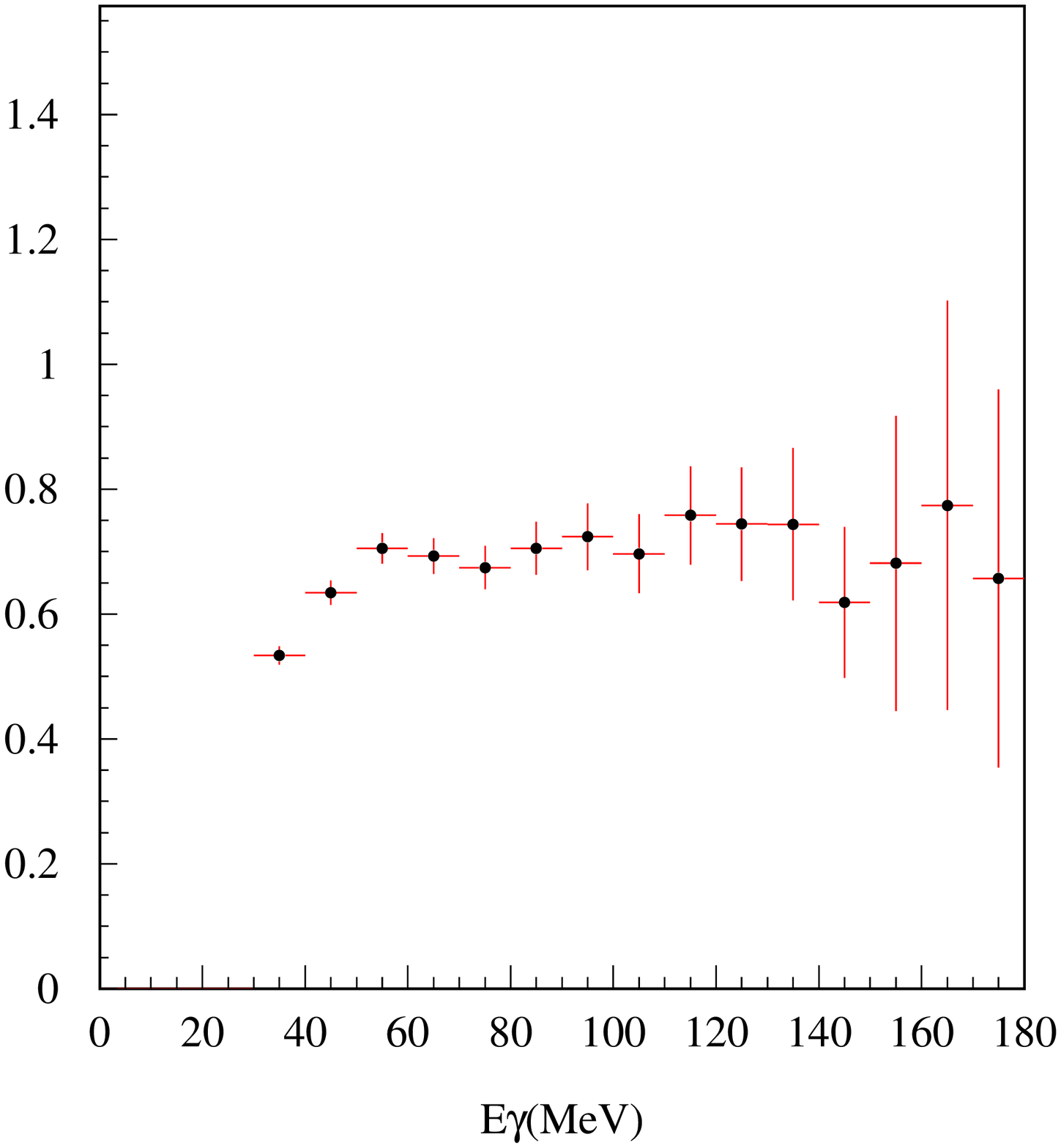,width=6.0cm} &
    \psfig{figure=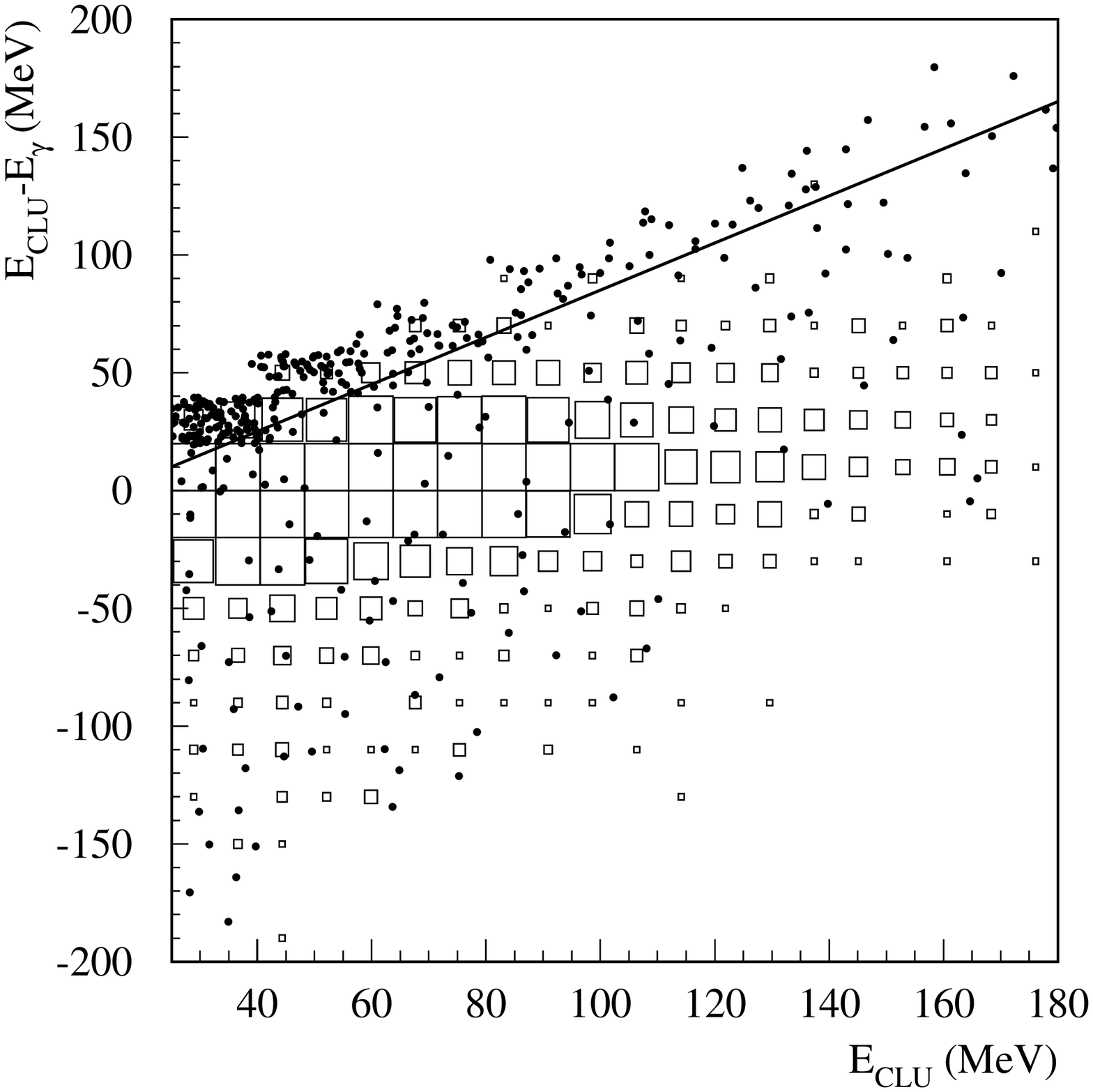,width=6.0cm}\\
    (a) & (b)
  \end{tabular}  
  \caption{Monte Carlo simulation: (a) \keg\ signal efficiency (one run period); (b) applied cut
  to remove accidentals.}
  \label{fig:effgsel}
\end{figure}

Fig.~\ref{fig:effgsel}(a) gives the selection efficiency for the signal as taken from Monte 
Carlo simulation.
The main background contribution comes from \keg\ events with an undetected 
soft-photon to which a cluster from accidentals has been wrongly associated.
This background is strongly reduced by requiring $E_{\rm{clu}}>$25~\mev\ and 
$E_{\rm{clu}}- E^{\rm{lab}}_{\gamma}< E_{\rm{clu}} - 15 $  \mev\, where $E_{\rm{clu}}$ is the energy of the 
photon cluster and $E^{\rm{lab}}_{\gamma}$ is the reconstructed photon energy in the laboratory 
system. This cut is shown in Fig.~\ref{fig:effgsel}(b). We obtain a factor 10 in background 
reduction with $\sim$6\% loss in signal efficiency.

Background from \klppp\ and \klpmn\ events after signal selection is at level of $\sim$ 4.2\% 
and $\sim$ 2.5\%, respectively. As shown in Fig.~\ref{fig:4plot}, this background overlaps with 
the signal of interest DE, so we need to remove it as much as possible.
To remove both \klppp\ and \klpmn\ we use a neural network (NN).
To remove \klppp\ events we use a NN based on the photon energy and angle (\wrt\ the lepton), 
the track momenta, the missing momentum and $M^2_{\gamma\nu}$, the invariant mass of 
photon-neutrino pair. 
To remove \klpmn\ events we use a NN based on the track momenta, the calorimetric energy 
deposit and the cluster centroid position.
Appropriate cuts on the NN output give a background reduction 
from 4.2\% $\to$ 0.4\% and from 2.5\% $\to$ 1.4\%, respectively for \klppp\ and \klpmn, with 
a signal loss of 10\%.
To check the data-MC agreement, to calibrate the MC position $\vec{X}_N$ and correct the 
photon selection efficiency we use \klppp\ decay events as a control sample. 
These events are selected using a tight kinematical cut in the variable 
$E^2_{\rm{miss}}-p^2_{\rm{miss}}-m^2_{\pi^0}$ in the hypothesis of two pion tracks.
Further, we require the presence of a cluster (E$>$60\mev) not associated to any track, 
corresponding to one of the two photons from $\pi^0$ decay. This high energy photon is used to 
tag the presence of the second photon. We select about 350,000 \klppp\ events with a purity of 
99.8\%.
\\
As a first check of Data-Monte Carlo agreement we compare the energy resolution of the
photons. This can be done because in this control sample we estimate the energy of the second 
photon (the tagged photon) exactly in the same way (Eq.~\ref{eq:enegcs}) than in our \keg\ 
signal selection: there we do not detect the neutrino, here we ignore the hard photon (the
tagging photon). Squaring the equation below the (second) photon energy is extracted.
\begin{equation}
  p_{\gamma-hard} = p_K - p_{\pi} - p_{\pi} - p_{\gamma} ~~~;
  ~~~ \pvec_{\gamma} = E_{\gamma}~\frac{\vec{X}_{\rm{clu}}-\vec{X}_N}{|\vec{X}_{\rm{clu}}-\vec{X}_N|}   
  \label{eq:enegcs}
\end{equation}
\begin{figure}[ht]
  \centering
  \begin{tabular}{cc}
    \psfig{figure=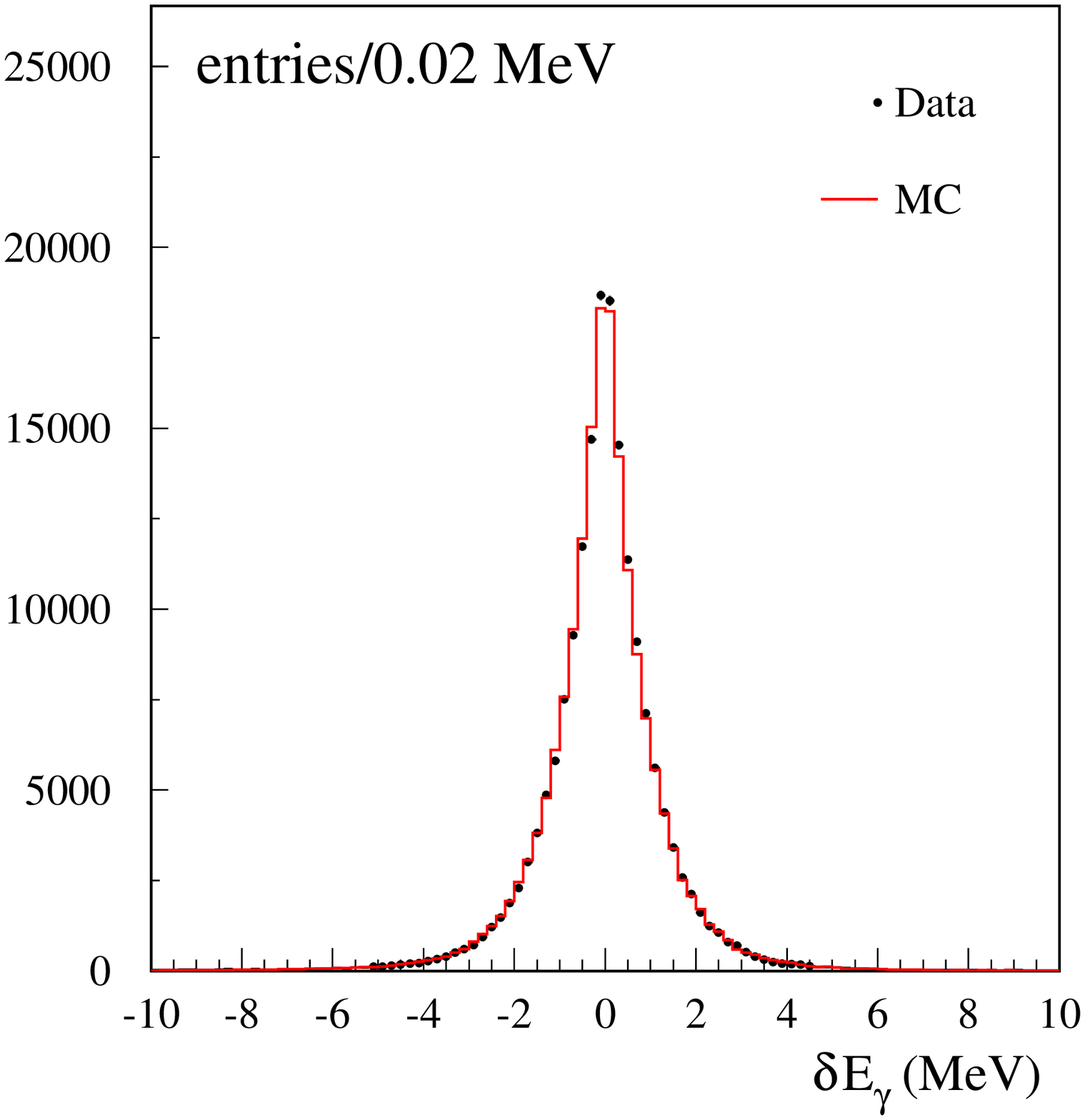,width=6cm}
    &
    \psfig{figure=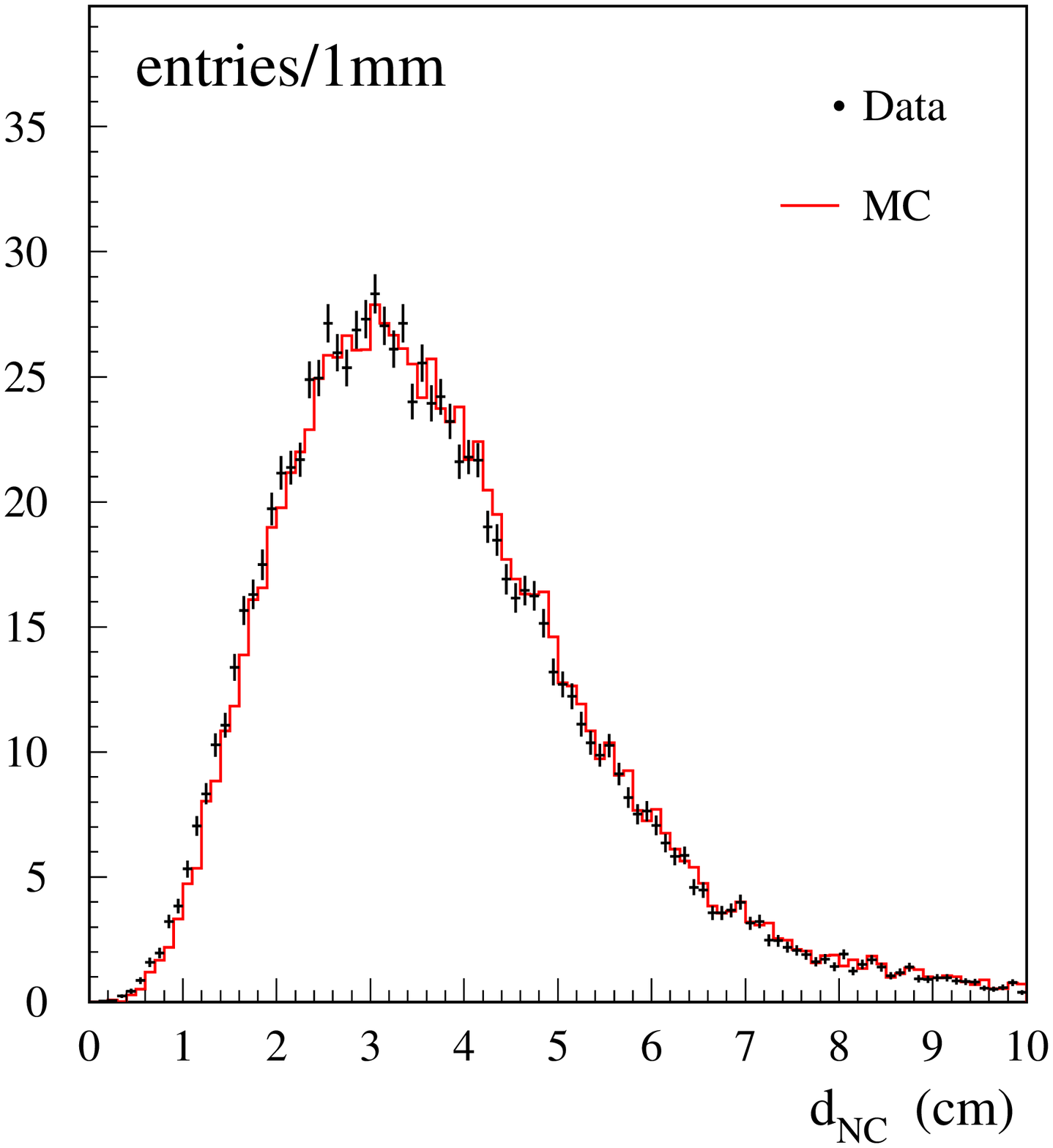,width=6cm}\\
  \end{tabular}
  \caption{From \klppp\ control sample: (a) photon energy resolution; 
    (b) $d_{NC}$ distribution in the central part of the drift chamber after correction.}
  \label{fig:gammacs}
\end{figure}
The photon energy resolution is evaluated with respect to a value computed in a
more accurate method using the complete hard photon informations and closing in this way the kinematic. 
In Fig.~\ref{fig:gammacs}(b) the residual is shown for data and Monte Carlo and a good agreement 
results.
\\
Further, we use the control sample from \klppp\ to evaluate the distance \dnc\ and its resolution $\sigma_{\dnc}$, 
in order to correct \dnc\ and $\sigma_{\dnc}$ in MC simulation.
Because of the use of $E_{\rm{clu}}$ to remove accidentals, we also use this control sample
to check the calorimeter energy response. The MC energy response is about $\sim$2\mev\ lower than data. 
To a good approximation, this bias is independent of the energy.
\\
Finally, we evaluate the photon selection efficiency from data and MC in this control sample and use 
their ratio to correct photon selection efficiency in MC simulation. The correction is of the order of 
a few percent.

\section{Fit} 
\label{sec:fit}
\begin{figure}[ht]
   \centering
   \psfig{figure=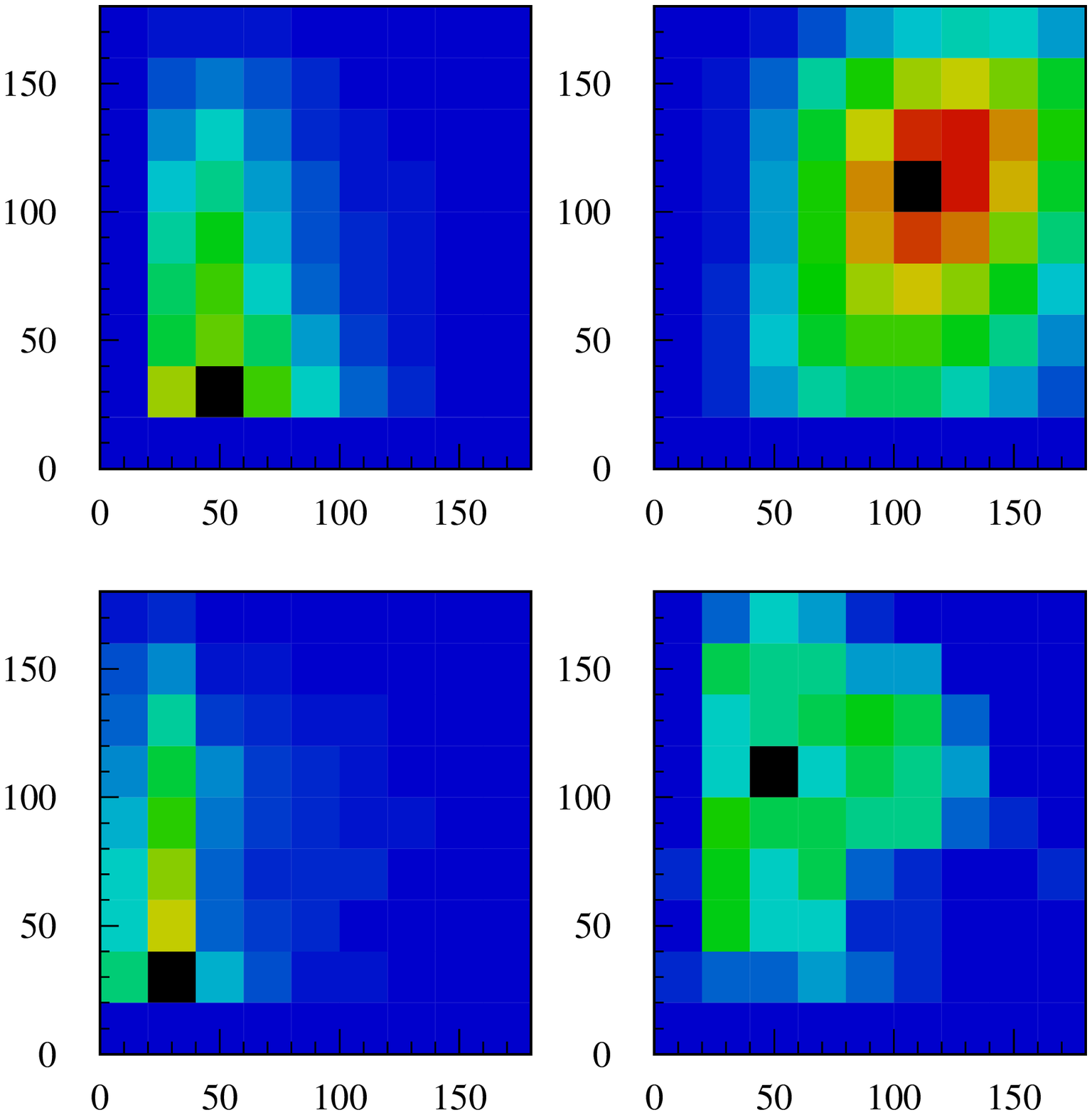,width=8cm}
   \caption{Monte Carlo distribution, \qstar\ (deg) \vs\ \Estar\ (\mev). From left-top: IB signal, 
     DE signal, \keg-out-of-acceptance and background from \klppp\ and \klpmn. 
     The vertical scale is in arbitrary units.}
   \label{fig:4plot}
 \end{figure}

\begin{figure}[ht]
  \centering
  \begin{tabular}{cc}
    \psfig{figure=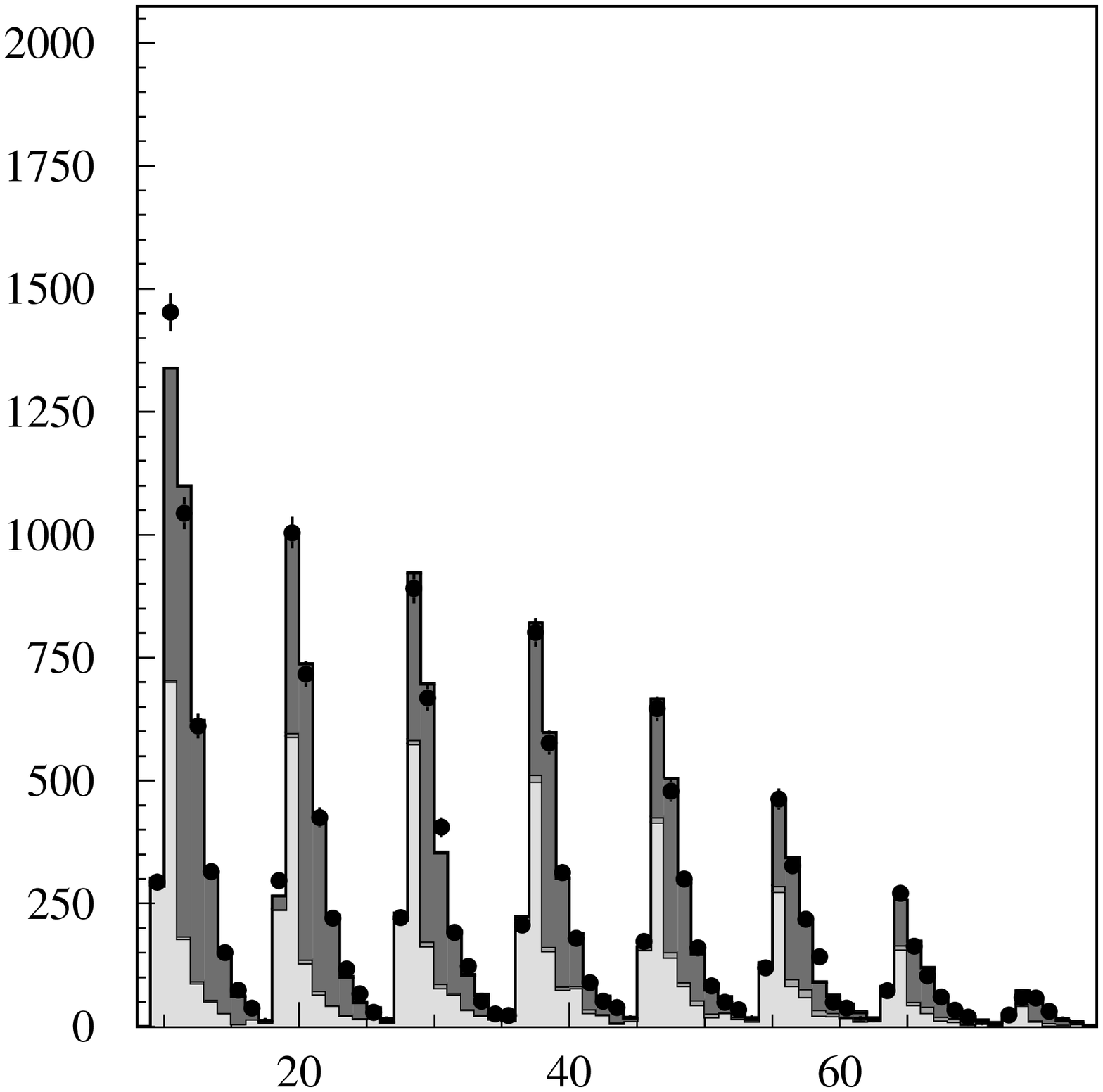,width=6.2cm}
    &
    \psfig{figure=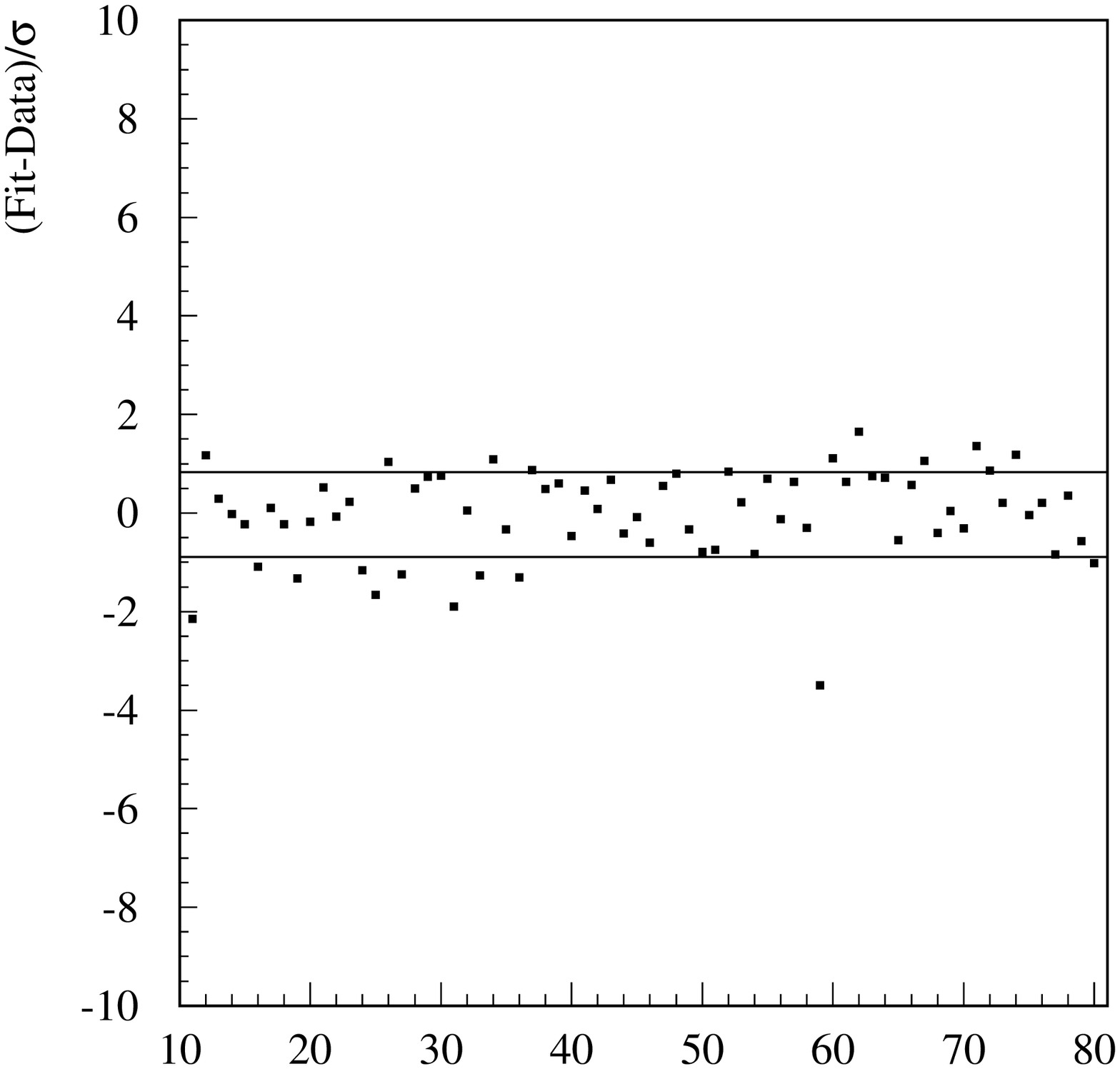,width=6.2cm}\\
    (a) & (b)\\
  \end{tabular}
  \caption{(a) Fit: dots are data, dark grey is the signal (IB+DE), light grey is the 
    \keg-out-of-acceptance; (b) Fit residual.}
  \label{fig:fit}
\end{figure}

In order to count \keg\ signal events we fit the Monte Carlo spectra $f_i(\Estar,\qstar)$ to the data 
($i=1,2,3,4$ respectively for IB signal, DE signal, \keg out-of-acceptance 
($\Estar<30\mev$ or $\qstar<20^{\circ}$) and physical background from \klppp\ and \klpmn\ events). 
The four distributions used as inputs in the fit are shown in Fig.~\ref{fig:4plot}.
Actually it is possible to measure R by using the energy spectrum of the photon only (we do it and the 
result matches), but in this case there is no sensitivity to the presence of a DE term. Only 
a simultaneous use of the energy \Estar\ and the angle \qstar\ can disentangle the small DE signal. 
To check the fit stability as a function of run period we fix the background component, otherwise
the fit could not converge (too low background statistic) and we do not use the DE shape (no 
sensitivity in a single run period). The stability is good ($\chi^2/\rm{dof}=9/13$). 
Then we fit all data simultaneously: free parameters of the fit are the population for IB signal,
DE signal, \keg-out-of-acceptance. We fix the background contribution of \klppp\ and \klpmn\ from 
MC.
As a check we also perform the fit letting all parameters free. The result matches well
but there is a loss in statistical accuracy. For this reason the background fraction 
is not a free parameter of the fit. 
The result of the fit and the residual are shown in Fig.~\ref{fig:fit}. The two-dimensional 
MC input shapes are arranged as 8 $\theta-slices$ one dimensional hystograms.  
Each {\it{slice}} covers 20 degrees, from 20$^{\circ}$ to 180$^{\circ}$.
The counting result and the correlations of parameters are given, respectively, in Table~
\ref{tab:count} and Table~\ref{tab:correl}. The $\chi^2$ is good, giving $\chi^2$/dof=60/69).
\begin{table}[hbt]
  \begin{center}
    \begin{tabular}{|c|c|c|c|}
      \hline 
      &   \multicolumn{3}{|c|}{ Counting result } \\
      \hline
                  & IB signal & \keg-out-acc & DE signal \\
      \hline
      counting    &  9083     &     6726       &  -102     \\
      error       &   213     &      194       &    59     \\  
      \hline
    \end{tabular}  
    \caption{Conting from the fit with the statistical error}
    \label{tab:count}    
  \end{center}
\end{table}
\begin{table}[hbt]
  \begin{center}
    \begin{tabular}{|c|c|c|c|}
      \hline 
      &   \multicolumn{3}{|c|}{ Correlation coefficients} \\
      \hline
      Par   &    1     &    2     &    3     \\
      \hline
      1    &  1.000   &  -0.586  &  -0.254   \\
      2    &          &   1.000  &  -0.022   \\  
      3    &          &          &   1.000   \\
      \hline
    \end{tabular}  
    \caption{Correlation coefficients: 1: IB signal, 2: \keg\ -out-of-acceptance, 3: DE signal}
    \label{tab:correl}    
  \end{center}
\end{table}
We get the \kegf\ events by counting events after \tof\ selection, bakground subtraction and efficiency 
correction. After getting \keg\ signal from fit counting, taking into account the efficieny and all
the efficiency corrections we measure:
\begin{equation}
  R = \frac{\br(\keg;\Estar>30\mev,\qstar>20^\circ)}{\br(\kegf)} = (924 \pm 23)\times 10^{-5}
\end{equation}
Although the $\chi^2$ for the fit is not bad if we use the \kloe\ MC to describe the IB spectrum
($\chi^2$ probability $\sim$55\%)  the {\it{Kubis}} Monte Carlo is slightly better, giving a 
77\% $\chi^2$ probability.

\section{Systematic uncertainties} 
We estimate all the systematics uncertainties by varying the cuts. Tracking, clustering, 
track-to-cluster association, NV acceptance and analysis cuts all depends on some parameters 
which define our signal. Any variation on these parameters produces a variation on the result.
In the following we list the absolute variation of $10^5\times$ R.\\
{\it{Tagging.}}
We tag the \kl\ requiring that \ks\ alone satisfies the 
calorimeter trigger with the presence of two clusters from \kspp\ associated with fired 
trigger sector ({\em{autotrigger}}). We observe a change of 4.\\
{\it{Tracking.}}
The most effective variable
in the tracking candidates definition is \dc , the distance of closest approach of the track to the 
tagging line. We vary \dc\ by a factor of two. We re-evaluate for each different configuration the 
tracking-efficiency correction,which is run-period dependent. The uncertainty on the tracking 
efficiency correction is dominated by sample statistics. We observe a change of 1.5.\\
{\it{Clustering.}}
The most effective variable in the definition of \tca\ association
is the transverse distance, \dtTCA.  We vary the cut on \dtTCA\ from 15 cm to 30 cm,
corresponding to a change in efficiency of about 17\%. 
We re-evaluate for each different configuration the clustering efficiency correction, which 
is run-period dependent. Also in this case the uncertainty on the clustering efficiency corrections 
is dominated by sample statistics.
We observe a change of 5.5 in the result.\\
{\it{Kinematic cuts.}}
We apply loose kinematic cuts. When varying this cut negligible variation for the results are found.\\
{\it{Tof cut.}}
Inclusive \kegf\ sample is identified also by using time of flight (\tof). We use a 2-$\sigma$
cut. After varying this cut by 30\% we observe a change of 1.3.\\
{\it{Momentum mis-calibration and resolution.}}
The effect of the momentum scale and the momentum resolution have also been considered. 
We conservatively assume a momentum scale uncertainty of 0.1\%
We observe a change of 3 for R.\\
We also investigate the effect of momentum resolution by changing its value of $\pm$ 3\%,
corresponding to a worst $\chi^2$. The variation on the result is 7.2.\\
{\it{Fiducial volume.}}
We reduce the fiducial volume by a 20\%. This produces a variation of 3.\\
{\it{Rejection of accidentals.}}
We vary the sliding cut used to remove wrong associations of accidental
cluster (see Fig.\ref{fig:effgsel}). Varying this cut we have a background variation of a factor of two.  
The change in the result is 5.2.\\
{\it{NV acceptance.}}
We search neutral vertex within a well defined sphere centered around $X_C$. We vary the
dimension of its radius by a factor of two. We observe a change of 2.9 in the result.\\
{\it{Background.}}
In a very conservatively way we remove the cut on NN output: in this way the background 
level increses almost of a factor of four. We observe a change of 9 for the result.\\
All systematic errors are summarized in Table \ref{tab:syst}.
\begin{table}[hbt]
  \begin{center}
    \begin{tabular}{|c|c|c|}
      \hline
      Source            &  $10^5\times\Delta R$ & $\Delta\langle X \rangle$ \\
      \hline
      Tagging           &    4.0   &  0.7   \\
      Tracking          &    1.5   &  0.8   \\  
      Clustering        &    5.5   &  0.1   \\
      Kinematic cuts    &  $\sim$0 & $\sim$0  \\
      TOF-cut           &    1.3   &  0.5   \\
      p-miscalibration  &    3.5   &  0.2   \\
      p-resolution      &    7.2   &  0.4   \\
      Fiducial volume   &    3.0   &  0.5   \\
      \hline
      Rejection acc.    &    5.2   &  0.4   \\
      NV acceptance     &    2.9   &  0.3   \\
      BKG Reject.by NN  &    9.0   &  0.1   \\
      \hline
      Total             &    15.5  &  1.4   \\
      \hline
    \end{tabular}  
    \caption{Summary of the absolute systematic uncertainties on R and 
      $\langle X \rangle$}
    \label{tab:syst}    
  \end{center}
\end{table}

\section{Results} 
Our final result for R is
\begin{equation}
  R = (924\pm 23_{\rm{stat}}\pm 16_{\rm{syst}})
  \times10^{-5}
\end{equation}
We also estimate the $\langle X \rangle$ parameter, defined in the equation~\ref{eq:x}.
Starting from counting result for IB and DE, taking into account the difference in the 
efficiency (IB efficiency is about $\sim$20\% higher than DE efficiency), including all systematics 
we measure
\begin{equation}
  \langle X\rangle_{meas} = -2.3\pm 1.3_{\rm{stat}}\pm1.4_{\rm{syst}} 
\end{equation}
in agreement with $\mathcal{O}(p^6)$ evaluation.
The systematics on $\langle X \rangle$ are evaluated as for R.
The different contributions are listed in Table~\ref{tab:syst}.
The presence of DE contribution reduces the value of R of about 1\%. The correlation between R 
and $\langle X \rangle$, including also systematics, is 3.9\%.
\begin{figure}[ht]
  \centering
   \psfig{figure=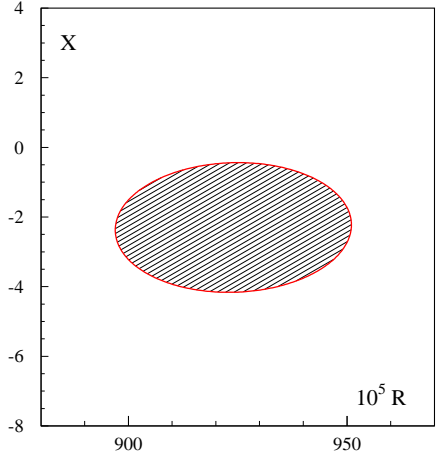,width=6cm}
   \caption{1-$\sigma$ confidence level for R and $\langle X \rangle$ measurement.} 
   \label{fig:correl}
 \end{figure}

\section{Conclusion} 
Two different components contribute to the photon emission in \keg\ events, the inner 
bremsstrahlung and the direct emission. The latter describes photon radiation from intermediate 
hadronic states, giving in this way new information on the hadronic structure of kaons.
Using \Estar\ and \qstar\ variables \kloe\ measured the width for \klpeng\ for $\Estar>30\mev$ and
$\qstar>20^\circ$ to the width for inclusive \kegf. 
The DE emission contribution originates from the interference with IB, resulting in a
negative effective strength, $\langle X \rangle$. \kloe\ measurement of $\langle X \rangle$ is the 
the first attempt to measure direct emission contribution in \keg\ process.
At this stage, the \kloe\ measurement of R (3\% accuracy) is not sufficient to solve the experimental 
disagreement between NA48 and KTeV measurement.

\section*{Acknowledgements} 
We would like to thank Bastian Kubis, one of the authors of Ref.\cite{kubis}, for the use of
his Monte Carlo generator in this analysis.
We thank the DAFNE team for their efforts in maintaining low background running conditions and 
their collaboration during all data-taking. 
We want to thank our technical staff: 
G.F.Fortugno for his dedicated work to ensure an efficient operation of 
the KLOE Computing Center; 
M.Anelli for his continuous support to the gas system and the safety of
the detector; 
A.Balla, M.Gatta, G.Corradi and G.Papalino for the maintenance of the
electronics;
M.Santoni, G.Paoluzzi and R.Rosellini for the general support to the
detector; 
C.Piscitelli for his help during major maintenance periods.
This work was supported in part by DOE grant DE-FG-02-97ER41027; 
by EURODAPHNE, contract FMRX-CT98-0169; 
by the German Federal Ministry of Education and Research (BMBF) contract 06-KA-957; 
by Graduiertenkolleg `H.E. Phys. and Part. Astrophys.' of Deutsche Forschungsgemeinschaft,
Contract No. GK 742; 
by INTAS, contracts 96-624, 99-37; 
and by the EU Integrated Infrastructure
Initiative HadronPhysics Project under contract number
RII3-CT-2004-506078.

\bibliographystyle{elsart-num}
\bibliography{paper}

\end{document}